\documentclass[prd,aps,showpacs,superscriptaddress,nofootinbib,preprint]{revtex4-1}
\usepackage{amssymb}
\usepackage{graphicx}
\def\slashchar#1{\setbox0=\hbox{$#1$}
   \dimen0=\wd0 \setbox1=\hbox{/} \dimen1=\wd1
   \ifdim\dimen0>\dimen1 \rlap{\hbox to \dimen0{\hfil/\hfil}} #1
   \else  \rlap{\hbox to \dimen1{\hfil$#1$\hfil}} / \fi}

\begin{document}
\title{ Nuclear medium effects in $\nu(\bar\nu)$-nucleus deep inelastic scattering}
\author{H. \surname{Haider}}
\affiliation{Department of Physics, Aligarh Muslim University, Aligarh-202 002, India}
\author{I. \surname{Ruiz Simo}}
\affiliation{Departamento de F\'\i sica Te\'orica and IFIC, Centro Mixto
Universidad de Valencia-CSIC, Institutos de Investigaci\'on de
Paterna, E-46071 Valencia, Spain}
\affiliation{Departamento de F\'isica At\'omica Molecular y Nuclear, Universidad de Granada, E-18071 Granada, Spain}
\author{M. Sajjad \surname{Athar}}
\email{sajathar@gmail.com}
\affiliation{Department of Physics, Aligarh Muslim University, Aligarh-202 002, India}
\author{M. J.  \surname{Vicente Vacas}}
\affiliation{Departamento de F\'\i sica Te\'orica and IFIC, Centro Mixto
Universidad de Valencia-CSIC, Institutos de Investigaci\'on de
Paterna, E-46071 Valencia, Spain}

\begin{abstract}
 We study the nuclear medium effects in the weak structure functions $F_2(x,Q^2)$ and $F_3(x,Q^2)$ in the deep inelastic neutrino/antineutrino reactions in nuclei.
We use a theoretical model for the nuclear spectral functions which incorporates the conventional nuclear effects, such as Fermi motion, binding and nucleon correlations.
 We also consider the pion and rho meson cloud contributions calculated from a microscopic model for meson-nucleus self-energies. The calculations have been performed 
 using relativistic nuclear spectral functions which include nucleon correlations. Our results are compared with the experimental data of NuTeV and CDHSW. 
\end{abstract}
\maketitle
\section{Introduction}
Nuclear medium effects in the deep inelastic scattering processes have been widely discussed after the measurement and comparison of iron and deuterium electromagnetic 
structure functions $F_2^N(x,Q^2)$ by the European Muon Collaboration at CERN using charged lepton beams~\cite{EMC_Aubert}.  
Thereafter studies, both theoretical as well as experimental, have been made in several nuclei. 
 Presently most of the information on nuclear medium effects comes from the charged lepton scattering data. The weak structure functions $F_2^N(x,Q^2)$ and $F_3^N(x,Q^2)$ have also been measured
using neutrino (antineutrino) beams~\cite{Allasia, Berge, Varvell, Oltman, Seligman, Sidorov1999, Fleming, Tzanov}. 
More experiments are planned to obtain data in the deep inelastic region using neutrino/antineutrino beams that will complement the information obtained from 
the charged lepton scattering. The nuclear effects for the weak structure functions $F_2^A(x,Q^2)$ and $F_3^A(x,Q^2)$ 
may be in general different. Moreover, the nuclear correction for the weak structure function $F_2^A(x,Q^2)$ may be different from that of the electromagnetic 
structure function $F_2^{EM, A}(x,Q^2)$. The precise measurement of deep inelastic scattering $\nu(\bar\nu)$ cross section is also important in providing
 global fits of the parton distribution functions (PDFs) and due to the fact that most of the $\nu(\bar\nu)$ experiments are being performed with nuclear targets, 
the nuclear effects should be properly accounted for before extracting the free nucleon parton distribution function. In the determination of electroweak parameters, a good knowledge of the nuclear 
medium effects is required.

Furthermore, with the confirmation of the neutrino oscillation hypothesis in the atmospheric as well as accelerator based experiments, the target is to determine precisely the parameters of the 
neutrino mass mixing 
matrix (PMNS matrix), particularly to get some information on mixing angle $\theta_{13}$ and CP-violating phase $\delta$, using long baseline neutrino experiments and getting neutrinos from factories
as well as $\beta$-beam sources~\cite{nufact}. Most of these experiments are in the few GeV energy region. These high intensity neutrino sources are aimed to reduce the statistical uncertainties. 
Recently more efforts have been made to understand the 
systematic uncertainties~\cite{nuint}.  This is because in the region of a few GeV, which is sensitive to the determination of the parameters of PMNS matrix, the cross sections have not been very well
 measured.
Due to this reason various experiments are going on or have been proposed and lots of theoretical studies have been recently made for understanding the nuclear medium effects. These theoretical studies are
 mainly done for the quasielastic and one pion production processes, and recently some work on the two pion production, nucleon knock out reaction, hyperon production and single kaon production 
has been performed. In the case of deep inelastic scattering process induced by weak interaction, there are very few calculations where the dynamical origin of the nuclear medium effects has been 
studied~\cite{Petti, Sajjad}. In some theoretical analysis, nuclear medium effects have been phenomenologically described in terms of a few parameters which are determined from fitting the experimental data of 
charged leptons and (anti)neutrino deep inelastic scattering from various nuclear targets~\cite{Hirai, hirai, eskola, Bodek1, Eskola, Kovarik, Kovarik1, Schienbein1}. 

MINER$\nu$A~\cite{minerva} is
 taking data using neutrinos 
from NuMI Lab., and their aim is to perform cross section measurement in the neutrino energy region of 1-20 GeV and with various nuclear targets like carbon, iron and lead. This will experimentally
 complement
the present theoretical understanding of nuclear medium effects. NuSOnG experiment~\cite{nusong, nusong1} has been proposed at Fermi lab
 to study the structure functions in the deep inelastic region using neutrino sources. 
NuTeV collaboration~\cite{Tzanov} has reported the results on weak charged and neutral current induced (anti)neutrino processes on an iron target in the deep inelastic region. NOMAD~\cite{nomad}
 is doing data analysis of their experimental
results and very soon going to report the results for the structure functions and cross sections in carbon target using neutrino beam. 

 In this paper, we  study nuclear medium  effects on the structure functions $F_2$($x$,$Q^2$) and $F_3$($x$,$Q^2$) in iron and carbon nuclear targets. We use a relativistic nucleon 
spectral function~\cite{FernandezdeCordoba:1991wf} to describe the momentum distribution of nucleons in the nucleus and 
define everything within a field theoretical approach where nucleon propagators are written in terms of this spectral function. The spectral function has been calculated using the 
Lehmann's representation for the relativistic nucleon propagator and nuclear many body theory is used for calculating it for an interacting Fermi sea in nuclear matter. A local density approximation is
then applied to translate these results to finite nuclei~\cite{Marco, Sajjad, Sajjad1}. We have assumed the Callan-Gross relationship for nuclear structure functions ${F_2}^A(x)$ and
 ${F_1}^A(x)$. The contributions of the pion and rho meson clouds  are taken into account in a many body field theoretical approach which is  
based on Refs.~\cite{Marco,GarciaRecio:1994cn}. We have taken into account target mass correction following Ref.~\cite{schienbein} which has significant effect at low $Q^2$, moderate and high 
Bjorken $x$. To take into account the shadowing effect which is important at low $Q^2$ and low x, and modulates the contribution of pion and rho cloud contributions, we have followed the works of 
Kulagin and Petti~\cite{Kulagin, Petti}. Since we have applied the present formalism at low $Q^2$ also, hence we have not assumed the Bjorken limit. Recently, we have applied the present formalism to study nuclear effects in the electromagnetic structure function $F_{2}(x, Q^2)$ in nuclei 
in the deep inelastic lepton nucleus scattering~\cite{Sajjad1}, and found that
the numerical results agree with recent JLab results where the data for the ratios $R_{F2}^A(x,Q^2)=\frac{2F_2^A(x,Q^2)}{AF_2^{Deuteron}(x,Q^2)}$ have been obtained~\cite{Seely} and also to the 
some of the earlier experiments performed with the heavier nuclear targets. Motivated by the success of the present
 formalism~\cite{Marco, Sajjad, Sajjad1, Marco2}, in this paper, we have studied the nuclear medium effects in weak structure functions $F_2$($x$,$Q^2$) and $F_3$($x$,$Q^2$) and compared our results with the 
experimental results of NuTeV and CDHSW. 
Furthermore,
 we have obtained the ratio of the structure functions $R_{F_i}^A(x,Q^2)=\frac{2F_i^A(x,Q^2)}{AF_i^{Deuteron}(x,Q^2)}$~(i=2,3) to see how they do compare with the ratio obtained earlier for 
the electromagnetic 
structure function. Using these $F_2$($x$,$Q^2$) and $F_3$($x$,$Q^2$) structure functions we have obtained the differential scattering cross sections in iron and carbon nuclear targets. 
The results in iron are compared
with the available experimental data and the results in carbon would be a good test of our model when NOMAD~\cite{nomad} results would come up.
 
The plan of the paper is as follows. In Sect.~\ref{Formalism} we introduce some basic formalism for lepton-nucleon scattering, in Sect.~\ref{Sec:Medium-effect} we analyse the different nuclear effects,
in Sect.~\ref{Sec:Results} we present the results of our calculations and compare them with the available experimental results. In Sect.~\ref{Sec:Conclusion} we conclude our findings.
\begin{figure}
\begin{center}
\includegraphics[width=6cm,height=6cm]{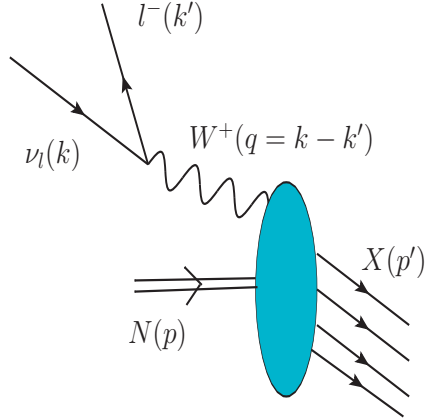}
\caption{Feynman diagram for the deep inelastic $\nu$-nucleon scattering}
\label{fg:fig1}
\end{center}
\end{figure}

\section{Formalism}\label{Formalism}
The expression of the differential cross section, for deep inelastic scattering (DIS) of neutrino with a nucleon target induced by charged current reaction 
\begin{equation} 	\label{reaction}
\nu_l(k) + N(p) \rightarrow l^-(k^\prime) + X(p^\prime),~l=~e,~\mu,
\end{equation}
shown in Fig.\ref{fg:fig1}, is given in terms of the Bjorken variables $x$ and $y$ and the dimensionless structure functions $F_i$(i=1-5) by
\begin{footnotesize}
\begin{eqnarray} \label{cross_section}
\frac{d^2\sigma^{\nu(\bar{\nu})}}{dx\ dy} &=& \frac{G_F^2 M
E_{\nu}}{\pi(1+Q^2/M_W^2)^2}\Biggl((y^2 x + \frac{m_l^2 y}{2 E_{\nu} M})
F_1(x,Q^2) +\left[ (1-\frac{m_l^2}{4 E_{\nu}^2})
-(1+\frac{M x}{2 E_{\nu}}) y\right]F_2(x,Q^2) \\ \nonumber
&
\pm&
\left[x y (1-\frac{y}{2})\right.
-\left.\frac{m_l^2 y}{4 E_{\nu} M}\right]
F_3(x,Q^2) 
+ \frac{m_l^2(m_l^2+Q^2)}{4 E_{\nu}^2 M^2 x} F_4(x,Q^2)
- \frac{m_l^2}{E_{\nu} M} F_5(x,Q^2)\Biggr)\
\end{eqnarray}
\end{footnotesize}
where $G_F$ is the Fermi coupling constant, $m_l$ is the mass of lepton, $E_\nu$ is the incident neutrino/antineutrino energy, M is the mass of nucleon, in $F_3$, +sign(-sign) is for neutrino(antineutrino), 
x($=\frac{Q^2}{2M\nu}$) is the Bjorken variable, $y=\frac{\nu}{E_\nu}$, $\nu$ and q being the energy
and momentum transfer of leptons and $Q^2=-q^2$. $F_4$ and $F_5$ are generally omitted since they are suppressed by a factor of at least
$m_l^2/2ME_\nu$ relative to the contributions of $F_1$, $F_2$ and
$F_3$. $F_1$ and $F_2$ are related by the Callan-Gross relation~\cite{Callan} leading to only
two independent structure functions  $F_2$ and $F_3$. The nucleon structure functions are determined in terms of parton distribution functions for quarks and anti-quarks. 
For the numerical calculations, parton distribution
 functions for the nucleons have been taken from the parametrization of CTEQ Collaboration (CTEQ6.6)~\cite{cteq}. The NLO evolution of the deep inelastic structure functions has been taken from the works of 
Moch et al.~~\cite{Vermaseren,Moch,Neerven}. 
\section{Nuclear effects in neutrino scattering}\label{Sec:Medium-effect}
When the reaction given by Eq.(\ref{reaction}) takes place in a nucleus several nuclear effects have to be considered. In general one may categorize these medium effects into two parts, 
a kinematic effect which arises as the struck nucleon is not at rest but is moving with a Fermi momentum in the rest frame of the nucleus and the other is a dynamic effect
which arises due to the strong interaction of the initial nucleon in the nuclear medium. 

In a nuclear medium the expression for the cross section is written as:
\begin{equation} 	\label{dif_cross_nucleus}
\frac{d^2 \sigma_{\nu,\bar\nu}^A}{d \Omega' d E'} 
= \frac{G_F^2}{(2\pi)^2} \; \frac{|{\bf k}^\prime|}{|{\bf k}|} \;
\left(\frac{m_W^2}{q^2-m_W^2}\right)^2
L^{\alpha \beta}_{\nu, \bar\nu}
\; W_{\alpha \beta}^A\,,
\end{equation}
where $W_{\alpha \beta}^A$ is the nuclear hadronic tensor defined in terms of nuclear hadronic structure functions $W^A_{i}(x,Q^2)$: 
\begin{eqnarray}\label{had_ten}
W^A_{\alpha \beta} &=&
\left( \frac{q_{\alpha} q_{\beta}}{q^2} - g_{\alpha \beta} \right) \;
W_1^A
+ \frac{1}{M_A^2}\left( p_{\alpha} - \frac{p . q}{q^2} \; q_{\alpha} \right)
\left( p_{\beta} - \frac{p . q}{q^2} \; q_{\beta} \right)
W_2^A \\ \nonumber
&-&\frac{i}{2 M_A^2} \epsilon_{\alpha \beta \rho \sigma} p^{\rho} q^{\sigma}
W_3^A
\end{eqnarray}
where $M_A$ is the mass of the nucleus. $L^{\alpha \beta}$ is the leptonic tensor given by
\begin{equation} 	\label{dif_cross2}
L^{\alpha \beta}=k^{\alpha}k'^{\beta}+k^{\beta}k'^{\alpha}
-k.k^\prime g^{\alpha \beta} \pm i \epsilon^{\alpha \beta \rho \sigma} k_{\rho} 
k'_{\sigma}\,,
\end{equation}
with - sign for neutrino and + sign for antineutrino in the antisymmetric term.

$W^A_i(x,Q^2)$ are redefined in terms of the dimensionless structure functions $F^A_{i}(x,Q^2)$ through 
\begin{eqnarray}\label{relation}
M_A W_1^A(\nu, Q^2)&=&F_1^A(x, Q^2) \\ \nonumber
\nu W_2^A(\nu, Q^2)&=&F_2^A(x, Q^2) \\ \nonumber
\nu W_3^A(\nu, Q^2)&=&F_3^A(x, Q^2)
\end{eqnarray}
In the local density approximation the reaction takes place at a 
point ${\bf r}$, lying inside the nuclear volume element $d^3r$ with local density $\rho_{p}({\bf r})$ and $\rho_{n}({\bf r})$ 
corresponding to the proton and neutron densities and the neutrino nuclear cross sections are obtained in terms of neutrino self energy $\Sigma(k)$ in the nuclear medium 
\begin{eqnarray}\label{Sigma}
\Sigma (k) =&& (-i)\frac{G_F}{\sqrt{2}}\frac{4}{m_\nu}
\int \frac{d^4 k^\prime}{(2 \pi)^4} \frac{1}{{k^\prime}^2-m_l^2+i\epsilon} 
\left(\frac{m_W }{q^2-m_W^2}\right)^2 \; L_{\alpha \beta} ~~ \Pi^{\alpha \beta} (q)\,.
\end{eqnarray}
where $\Pi^{\alpha \beta} (q)$ is the $W$ self-energy in the nuclear medium~\cite{Marco}:
\begin{eqnarray}\label{Self_1}
-i\Pi^{\alpha \beta} (q) =&&
(-) \; \int \frac{d^4 p}{(2 \pi)^4} iG(p) \;  \;
\sum_X \; \sum_{s_p, s_i} \prod^n_{i = 1}
\int \frac{d^4 p'_i}{(2 \pi)^4}
\prod_l i G_l (p'_l) \prod_j \; i D_j (p'_j) 
\left( \frac{-G_F m_W^2}{\sqrt{2}} \right)\nonumber\\
&&\times
\langle X | J^{\alpha} | N \rangle \langle X | J^{\beta} | N \rangle^*
(2 \pi)^4 \delta^4 (q + p - \Sigma^n_{i = 1} p'_i)\,.
\end{eqnarray}
X is the final state which consists of fermions and 
bosons. l and j are indices for the fermions and bosons respectively. $G_l (p'_l)$ and $D_j (p'_j)$ are respectively the nucleon and meson relativistic propagators in the final state~\cite{Itzykson}. 
G(p) is the nucleon propagator with mass M and energy E({$\bf p$}) in the initial state, which is calculated for a relativistic nucleon in the interacting Fermi sea 
by making a perturbative expansion of $G(p)$ in terms of $G^{0}(p)$, the free nucleon propagator, 
 and applying the ladder approximation to give~\cite{Marco}:
\begin{eqnarray}
G(p)=\frac{M}{E({\bf p})}\sum_{r}\frac{u_{r}({\bf p})\bar u_{r}({\bf p})}{p^{0}-E({\bf p})-\bar u_{r}({\bf p})\sum^N(p^{0},{\bf p})u_{r}({\bf p})\frac{M}{{\bf E(p)}}}
\end{eqnarray}
$u_{r}({\bf p})$ is the Dirac spinor with the normalization ${\bar u_{r}({\bf p})}u_{r}({\bf p})=1$ and $\Sigma^N(p^0,p)$ is the nucleon self energy in nuclear matter taken 
from Ref.~\cite{FernandezdeCordoba:1991wf}. 

The relativistic nucleon propagator G(p) in a nuclear medium is then expressed as~\cite{Marco}:
\begin{eqnarray}\label{Gp}
G (p) = \frac{M}{E({\bf p})} 
\sum_r u_r ({\bf p}) \bar{u}_r({\bf p})
\left[\int^{\mu}_{- \infty} d \, \omega 
\frac{S_h (\omega, \mathbf{p})}{p^0 - \omega - i \eta}
+ \int^{\infty}_{\mu} d \, \omega 
\frac{S_p (\omega, \mathbf{p})}{p^0 - \omega + i \eta}\right]\,,
\end{eqnarray}
 where $S_h (\omega, \mathbf{p})$ and $S_p (\omega, \mathbf{p})$ are the hole
and particle spectral functions respectively, $\mu$ is the chemical potential and for the present numerical calculations have been taken from Ref.~\cite{FernandezdeCordoba:1991wf}. 
We ensure that the spectral function is properly normalized and we get the 
correct Baryon number and binding energy for the nucleus. 

The cross section for neutrino scattering from an element of volume $d^3 r$ in the nucleus is given by~\cite{Sajjad}:
\begin{eqnarray}\label{change}
d \sigma & = & - \frac{2 m_{\nu}}{|{\bf k}|} \;
\hbox{Im} \; \Sigma \; d^3 r \,.
\end{eqnarray}
Using Eq.(\ref{Sigma}) in Eq.(\ref{change}), and using Eq.(\ref{dif_cross_nucleus}) we get the expression for the differential scattering cross section in the local 
density approximation with the hadronic tensor $W^A_{\alpha \beta}$
\begin{equation}	\label{conv_WA}
W^A_{\alpha \beta} = 4 \int \, d^3 r \, \int \frac{d^3 p}{(2 \pi)^3} \, 
 \int^{\mu}_{- \infty} d p^0 \frac{M}{E ({\bf p})} S_h (p^0, {\bf p}, \rho(r))
W^N_{\alpha \beta} (p, q), \,.
\end{equation}
where $W^N_{\alpha \beta} (p, q)$ is the hadronic tensor for the free nucleon target that is given by Eq.(\ref{had_ten}) with $M_A$ replaced by the mass of nucleon $M$.  

Using Eqs.(\ref{dif_cross_nucleus}), (\ref{had_ten}), (\ref{relation}) and (\ref{conv_WA}), we get the expressions for $F^A_2(x)$ and $F^A_3(x)$ as~\cite{Sajjad}:
\begin{eqnarray}\label{f2Anuclei}
F^A_2(x_A,Q^2)&=&4\int d^3r\int \frac{d^3p}{(2\pi)^3}\frac{M}{E({\bf p})}\int_{-\infty}^\mu dp^0\;
S_h(p^0,\mathbf{p},\rho(\mathbf{r})) \frac{x}{x_N}
\left( 1+\frac{2x_N p_x^2}{M\nu_N} \right)  F_2^N(x_N,Q^2)~~~~~
\end{eqnarray}
\begin{eqnarray}\label{f3Anuclei}
F_3^A(x_A,Q^2)&=&4\int d^3r \int \frac{d^3p}{(2\pi)^3} \frac{M}{E({\bf p})}\int_{-\infty}^{\mu} dp^0
S_h(p^0,\mathbf{p},\rho(\mathbf{r})) \frac{p^0\gamma-p_z}{(p^0-p_z\gamma)\gamma} F_3^N(x_N,Q^2)
\end{eqnarray}
where 
\begin{equation}	\label{gamma}
\gamma=\frac{q_z}{q^0}=
\left(1+\frac{4M^2x^2}{Q^2}\right)^{1/2}\,,
\end{equation}
and \[x_N=\frac{Q^2}{2(p^0q^0-p_zq_z)}\]
\subsection{$\pi$ and $\rho$ mesons contribution to the nuclear structure function}\label{Pion_Contribution}
\label{sec:meson}
The pion and rho meson cloud contributions to the $F_2$ structure function have been implemented following the many body field theoretical approach of Refs.~\cite{Marco,GarciaRecio:1994cn}. We have performed the numerical
calculations for isoscalar nuclear targets as the experimental results reported by the CDHSW\cite{Berge} and NuTeV~\cite{Tzanov} collaborations are corrected for the non-isoscalarity in the iron target. 
In the case of $F_3$ structure function there are no contributions from pion and rho meson clouds as it only gets contribution from valence quark distributions ($(u - \bar u) + (d - \bar d)$).
   
 The pion structure function $F_{2 A, \pi} (x)$ is written as~\cite{Marco};
\begin{equation}  \label{F2pion}
F_{2, \pi}^A (x) = - 6 \int  d^3 r  \int  \frac{d^4 p}{(2 \pi)^4} \; 
\theta (p^0) \; \delta I m D (p) \; 
\; \frac{x}{x_\pi} \; 2 M \; F_{2 \pi} (x_\pi) \; \theta (x_\pi - x) \; 
\theta (1 - x_\pi) 
\end{equation}
where $D (p)$ is the pion propagator in the nuclear medium which is given in terms of the pion self energy $\Pi_{\pi}$:
\begin{equation}
D (p) = [ p_0^{2} - \vec{p}\,^{2} - m^2_{\pi} - \Pi_{\pi} (p^0, p) ]^{- 1}\,,
\end{equation}
where
\begin{equation}\label{pionSelfenergy}
\Pi_\pi=\frac{f^2/m_\pi^2 F^2(p)\vec{p}\,^{2}\Pi^*}{1-f^2/m_\pi^2 V'_L\Pi^*}\,.
\end{equation}
Here, $F(p)=(\Lambda^2-m_\pi^2)/(\Lambda^2+\vec{p}\,^{2})$ is the $\pi NN$ form factor and  $\Lambda$=1 GeV, $f=1.01$, $V'_L$ is
the longitudinal part of the spin-isospin nucleon-nucleon interaction and $\Pi^*$ is the irreducible pion self energy that contains the contribution of particle - hole and delta - hole excitations. 
In Eq.(\ref{F2pion}), $\delta Im D(p)$ is given by  
\begin{equation}
\delta I m D (p) \equiv I m D (p) - \rho \;
\frac{\partial Im D (p)}{\partial \rho} \left|_{\rho = 0} \right.
\end{equation}
and
\begin{equation} 
\frac{x}{x_{\pi}} = \frac{- p^0 + p^z}{M}
\end{equation}
Assuming SU(3) symmetry and following the same notation as in Ref.\cite{Gluck:1991ey},  the pion structure function at LO can be written in terms of pionic PDFs as
\begin{equation}\label{pion_structure_function}
F_{2\pi}(x_\pi)=x_\pi\left(2v_\pi(x_\pi)+6\bar{q}_\pi(x_\pi)\right)
\end{equation}
where $v_\pi(x_\pi)$ is the valence distribution and $\bar{q}_\pi(x_\pi)$ is the light SU(3)-symmetric sea distribution.

Similarly, the contribution of the $\rho$-meson cloud to the structure function  is written as~\cite{Marco}
\begin{equation} \label{F2rho}
F_{2, \rho}^A (x) = - 12 \int d^3 r \int \frac{d^4 p}{(2 \pi)^4}
\theta (p^0) \delta Im D_{\rho} (p) \frac{x}{x_{\rho}} \, 2 M
F_{2 \rho} (x_{\rho}) \theta (x_{\rho} - x) \theta (1 - x_{\rho})
\end{equation}
\noindent
where $D_{\rho} (p)$ is the $\rho$-meson propagator and $F_{2 \rho}
(x_{\rho})$ is the $\rho$-meson structure function, which we have taken equal to the pion structure function $F_{2\pi}$ using the valence and sea pionic PDFs from reference \cite{Gluck:1991ey}. 
$\Lambda_\rho$ in $\rho NN$ form factor $F(p)=(\Lambda_\rho^2-m_\rho^2)/(\Lambda_\rho^2+\vec{p}\,^{2})$ has also 
been taken as 1 GeV.
In the case of pions 
we have taken the pionic parton distribution functions given by Gluck et al.~\cite{Gluck:1991ey,Gluck}.  For the rho mesons, we have applied the same PDFs as for the pions as in Refs.~\cite{Marco, Sajjad1}.
This model for the pion and $\rho$ self energies has been earlier applied successfully in the intermediate energy region and provides quite a solid description of a wide range of phenomenology  
in pion, electron and photon induced reactions in nuclei, see e.g. Refs. \cite{Marco,Sajjad1,Oset:1981ih,Carrasco:1989vq,Nieves:1991ye,Nieves:1993ev,Gil:1997bm}.
\subsection{Target mass corrections}
Target mass corrections have been incorporated by means of the approximate formula~\cite{schienbein}, which for $F_2^{\rm TMC}(x,Q^2)$ is given by
\begin{equation}\label{f2TMC}
 F_{2}^{TMC}(x,Q^2)\simeq\frac{x^2}{\xi^2\,\gamma^3} F_{2}(\xi,Q^2)\left[ 1+\frac{6\, \mu\, x\, \xi}{\gamma}(1-\xi)^2\right],
\end{equation}
and for $F_3^{\rm TMC}(x,Q^2)$ is given by
\begin{equation}
F_{3}^{{\rm TMC}}(x,Q^{2}) \simeq \frac{x}{\xi \gamma^{2}} F_{3}(\xi,Q^2)
\bigg[1-\frac{\mu x \xi}{\gamma} (1-\xi) \ln \xi \bigg]\, .
\label{eq:f3tmc_approx}
\end{equation}
where $\mu=\frac{M^2}{Q^2}$, $ \gamma = \sqrt{ 1 + \frac{4 x^2 M^2}{ Q^2 } }$ and $\xi$ is the Natchmann variable defined as
\begin{equation}\label{Natchmann}
\xi = \frac{2 x}{1+\gamma}\,.
\end{equation}
\begin{figure}
\includegraphics[width=16cm,height=12cm]{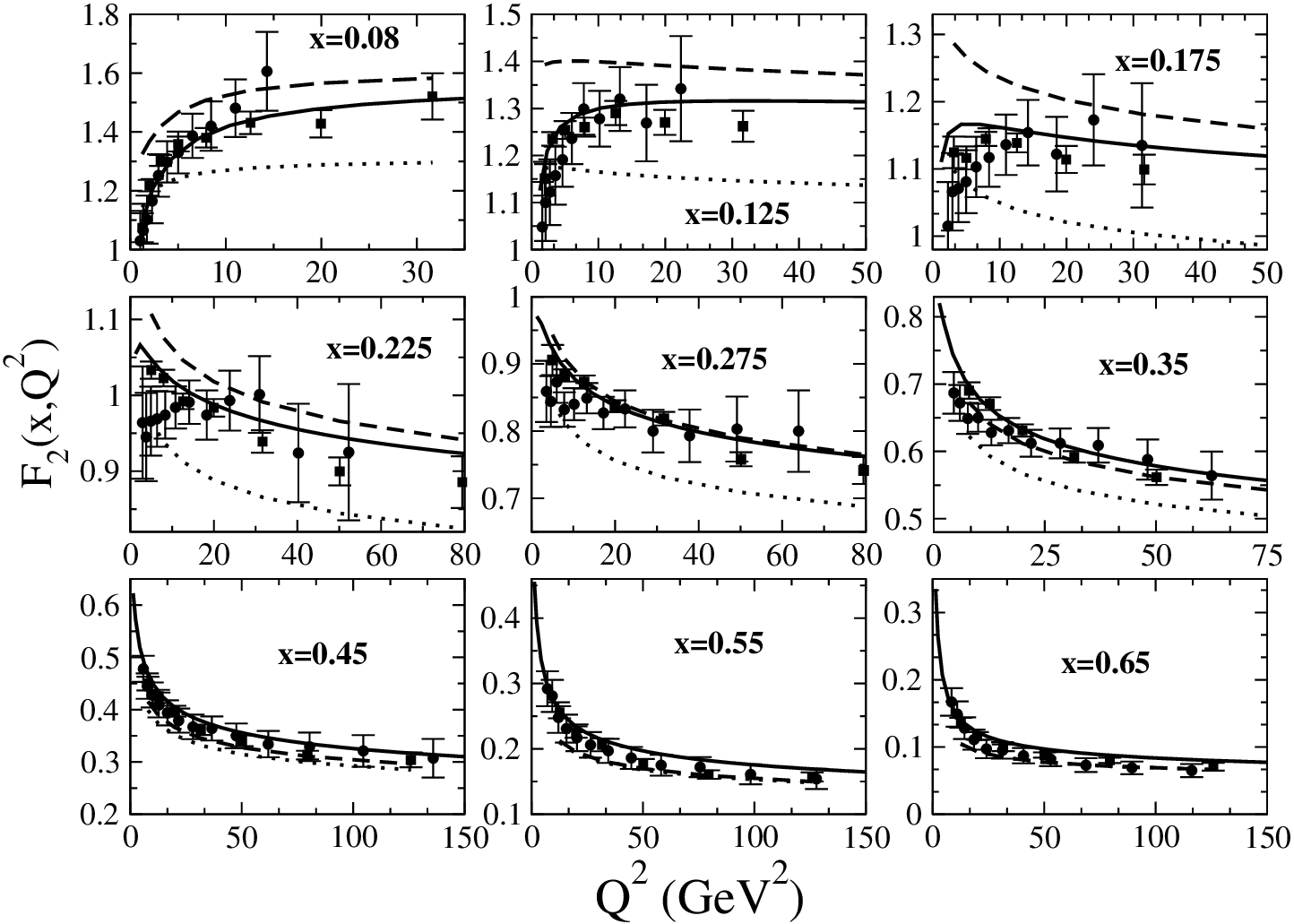}
\caption{Dotted line is $F_2(x,Q^2)$ vs $Q^2$ in $^{56}$Fe calculated using Eq.(\ref{f2Anuclei}) with TMC. For the calculations CTEQ~\cite{cteq} PDFs at LO have been used. 
Dashed line is the full model at LO. Solid line is full calculation at NLO. The experimental points are from CDHSW~\cite{Berge}(solid circle) and NuTeV~\cite{Tzanov}(solid square).}
\label{fig:f2}
\end{figure}
\begin{figure}
\includegraphics[width=16cm,height=12cm]{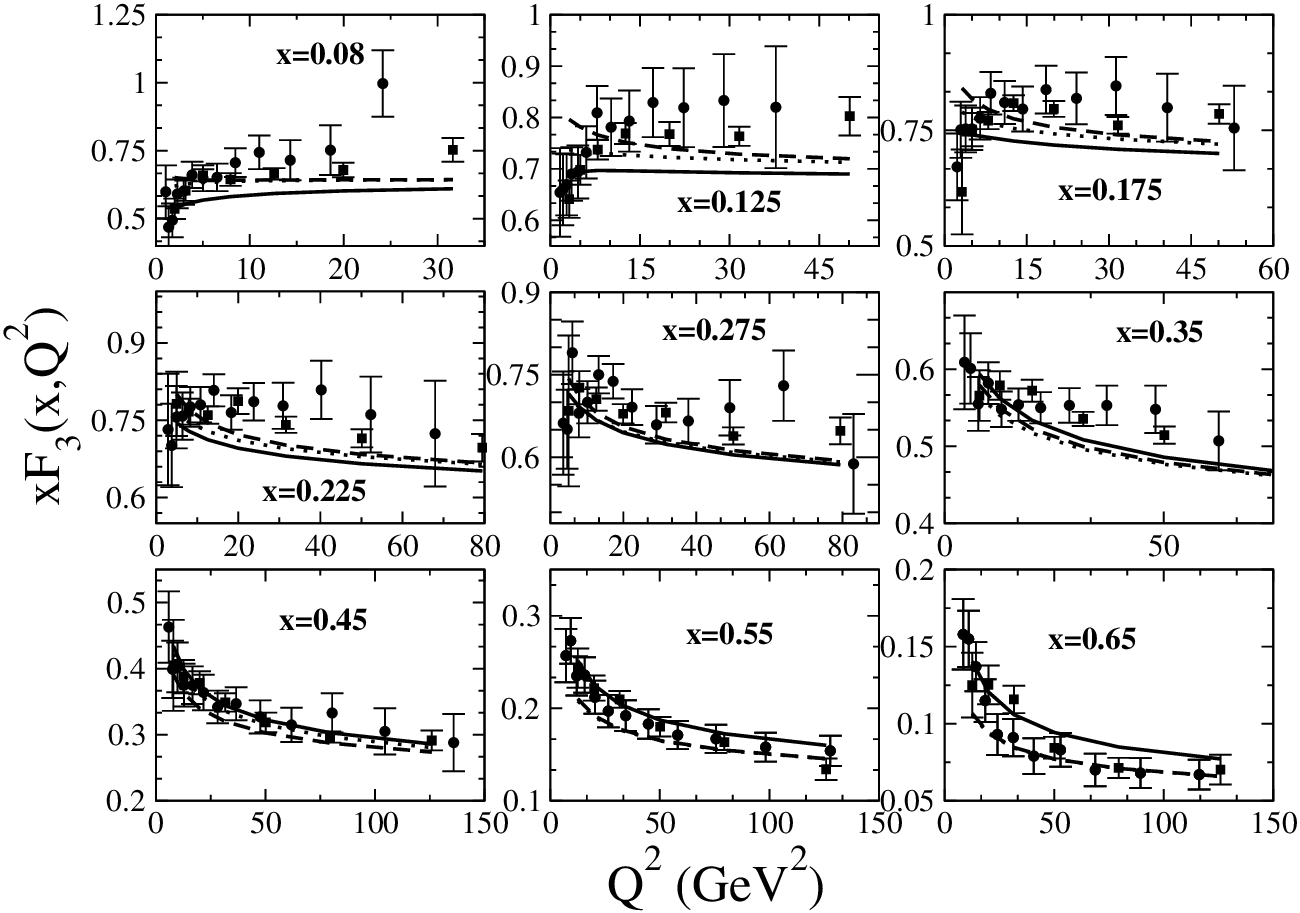}
\caption{Dotted line is $xF_3(x,Q^2)$ vs $Q^2$ in $^{56}$Fe calculated using Eq.(\ref{f3Anuclei}) with TMC. For the calculations CTEQ~\cite{cteq} PDFs at LO have been used. Dashed line is the full model at LO. 
Solid line is full calculation at NLO. The experimental points are from CDHSW~\cite{Berge}(solid circle) and NuTeV~\cite{Tzanov}(solid square).}
\label{fig:f3}
\end{figure}
\subsection{Coherent nuclear effects}\label{shadowing}
For the shadowing and antishadowing nuclear effects we use the model developed by Kulagin and Petti in Ref. \cite{Petti}.
We quote their formulae here only for completeness. Following their notation, we have the ratios for the
coherent nuclear correction to the structure functions $F_T$ and $F_3$, i.e, 
$\delta F^{\nu(\bar{\nu})A}_i=\delta R^{\nu(\bar{\nu})}_i F^{\nu(\bar{\nu})N}_i$ with $i=T,2,3$.
\begin{eqnarray}
\delta R^{\nu(\bar{\nu})}_T&=&\delta R^{(0,+)}\pm \beta\delta R^{(1,-)}\frac{F^{(\nu-\bar{\nu})(1)}_T}{2F_T^{\nu(\bar{\nu})N}}\\
\delta R^{\nu(\bar{\nu})}_3&=&\delta R^{(0,-)}\pm \beta\delta R^{(1,+)}\frac{F^{(\nu-\bar{\nu})(1)}_3}{2F_3^{\nu(\bar{\nu})N}}\pm(\delta R^{(0,+)}-
\delta R^{(0,-)})\frac{F_3^{(\nu-\bar{\nu})(s)}}{2F_3^{\nu(\bar{\nu})N}}
\end{eqnarray}
In the above equations, the labels $(I,C)$ with $I=0,1$ and $C=\pm$ stand for the classification in terms of isospin and C-parity of
the scattering states.
The parameter $\beta=\frac{Z-N}{A}$ must be set equal to $0$ if we are considering an isoscalar nucleus. Even for Fe-56 because we are considering it as an isoscalar nucleus. 
In the above equations the plus (minus) sign refers to neutrino (antineutrino). We assume the same correction factor for $F_2$ and 
$F_T$.
\begin{equation}
\delta F_2^{\nu(\bar{\nu})A}=\delta F_T^{\nu(\bar{\nu})A}=\delta R_T^{\nu(\bar{\nu})}F_T^{\nu(\bar{\nu})N}=\delta R_T^{\nu(\bar{\nu})}F_2^{\nu(\bar{\nu})N}
\end{equation}
\begin{figure}
\includegraphics[width=6cm,height=6cm]{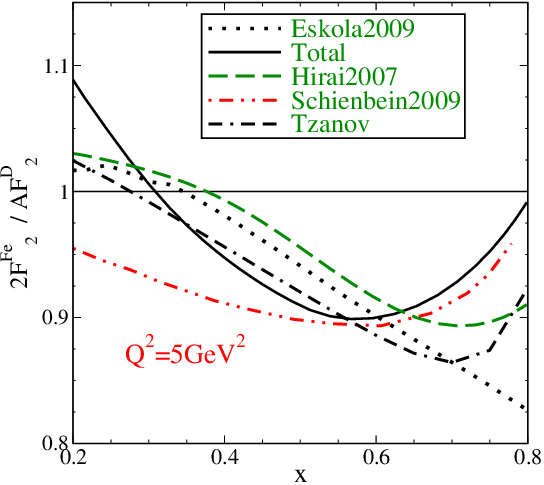}
\caption{Ratio R(x,$Q^2$)=$\frac{2F_{2}^{Fe}}{AF_2^D}$ with full calculation has been shown by the solid line.  
 Calculations have been done for $Q^2=5$ GeV$^2$ using CTEQ~\cite{cteq} PDFs at NLO. The results from Tzanov et al.~\cite{Tzanov}(double dashed-dotted), 
Hirai et al.~\cite{hirai}(dashed line), 
Eskola et al.~\cite{Eskola}(dotted line) and Schienbein et al.~\cite{Schienbein1}(double dotted-dashed line) have also been shown. }
\label{fig:f2ratio}
\end{figure}
\begin{figure}
\includegraphics[width=6cm,height=6cm]{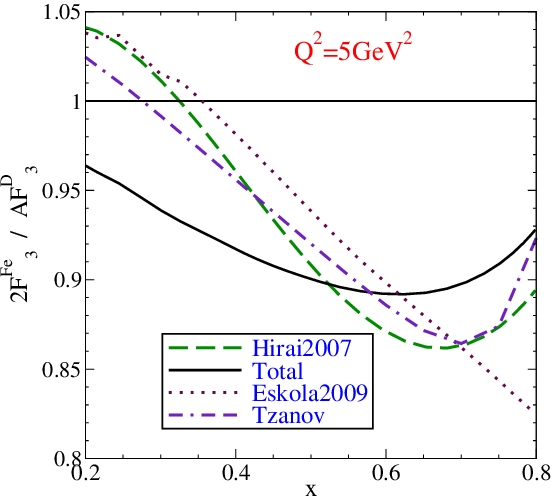}
\caption{Ratio R(x,$Q^2$)=$\frac{2F_{3}^{Fe}}{AF_3^D}$ with full calculation has been shown by the solid line.   
 Calculations have been done for $Q^2=5$ GeV$^2$ using CTEQ~\cite{cteq} PDFs at NLO. The results from Tzanov et al.~\cite{Tzanov}(double dashed-dotted), 
Hirai et al.~\cite{hirai}(dashed line) and
Eskola et al.~\cite{Eskola}(dotted line) have also been shown.}
\label{fig:f3ratio}
\end{figure}
\begin{figure}
\includegraphics[width=16cm,height=12cm]{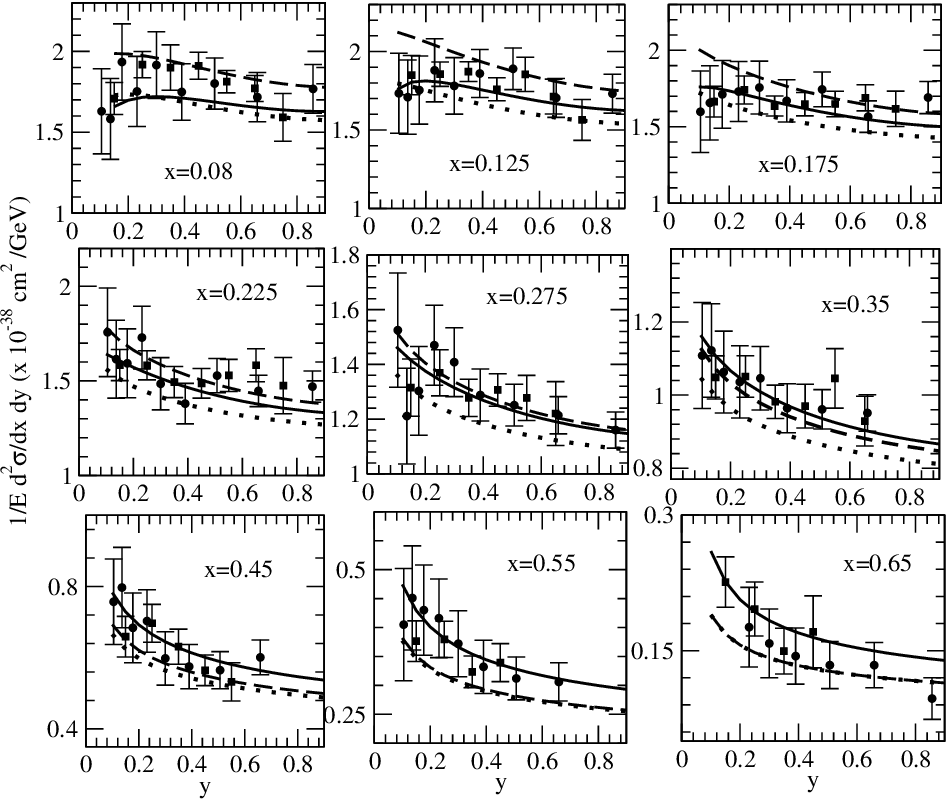}
\caption{$\frac{1}{E}\frac{d^2\sigma}{dxdy}$ vs y at different x for $\nu_\mu$($E_{\nu_\mu}=65$ GeV) induced reaction in $^{56}$Fe.
 Dotted (Dashed) line is the base (total) calculation at LO. Solid line is the full calculation at NLO.
 NuTeV~\cite{Tzanov} data have been shown by the squares and CDHSW~\cite{Berge} data by the solid circles. }
\label{neutrinoiron65gev}
\end{figure}
\begin{figure}
\includegraphics[width=16cm,height=12cm]{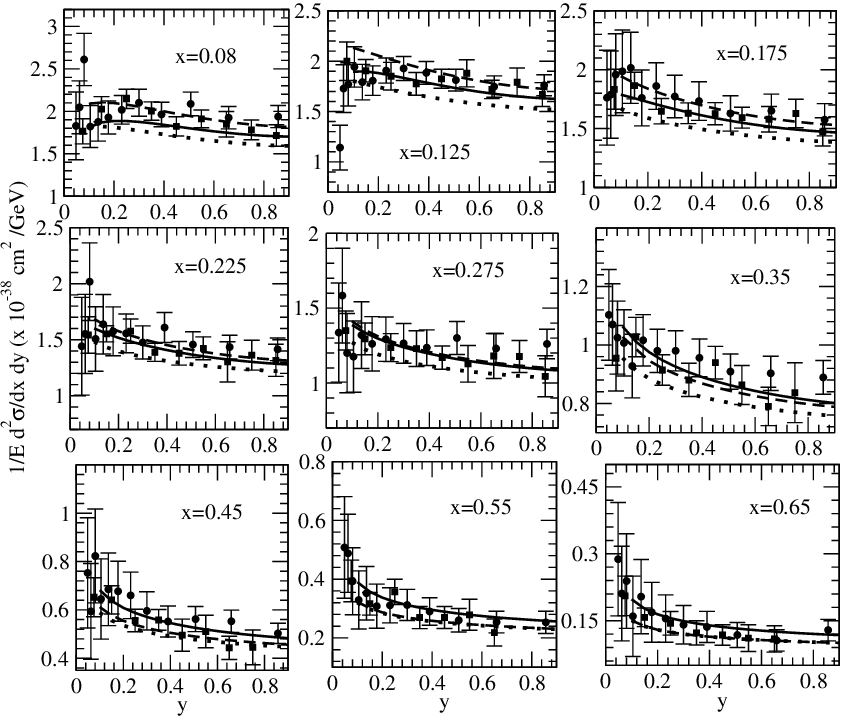}
\caption{$\frac{1}{E}\frac{d^2\sigma}{dxdy}$ vs y at different x for $\nu_\mu$($E_{\nu_\mu}=150$ GeV) induced reaction in $^{56}$Fe. Dotted (Dashed)
 line is the base (total) calculation at LO. Solid line is the full calculation at NLO. NuTeV~\cite{Tzanov} data have been shown by the squares and CDHSW~\cite{Berge} data by the solid circles. }
\label{neutrinoiron150gev}
\end{figure}
\begin{figure}
\includegraphics[width=16cm,height=12cm]{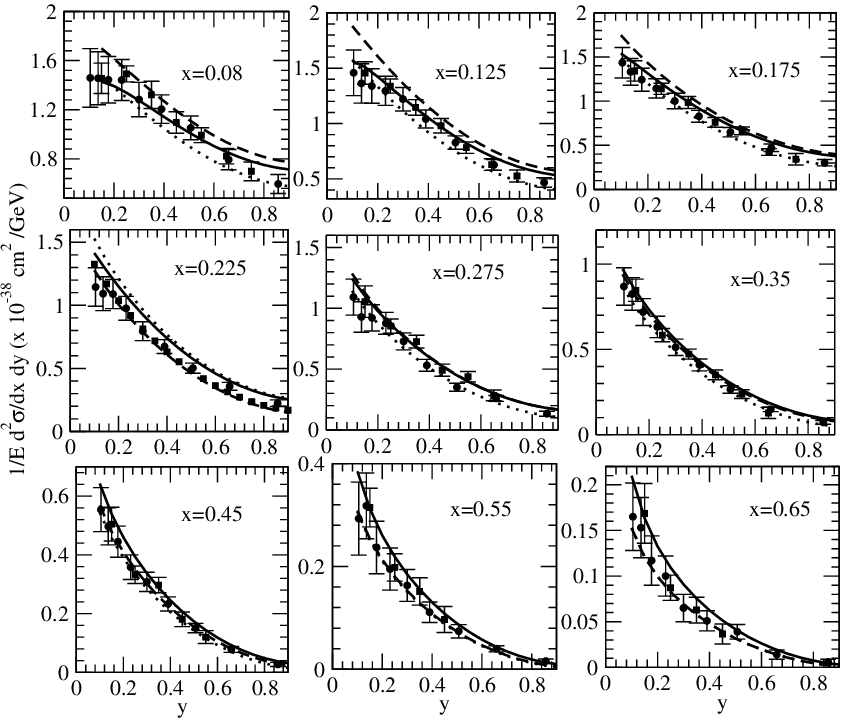}
\caption{$\frac{1}{E}\frac{d^2\sigma}{dxdy}$ vs y at different x for $\bar\nu{_\mu}$($E_{\bar\nu{_\mu}}=65$ GeV) induced reaction in $^{56}$Fe.
 Dotted (Dashed) line is the base (total) calculation at LO. Solid line is the full calculation at NLO. NuTeV~\cite{Tzanov} data have been shown by the squares and CDHSW~\cite{Berge} data by the solid circles. }
\label{antineutrinoiron65gev}
\end{figure}
\begin{figure}
\includegraphics[width=16cm,height=12cm]{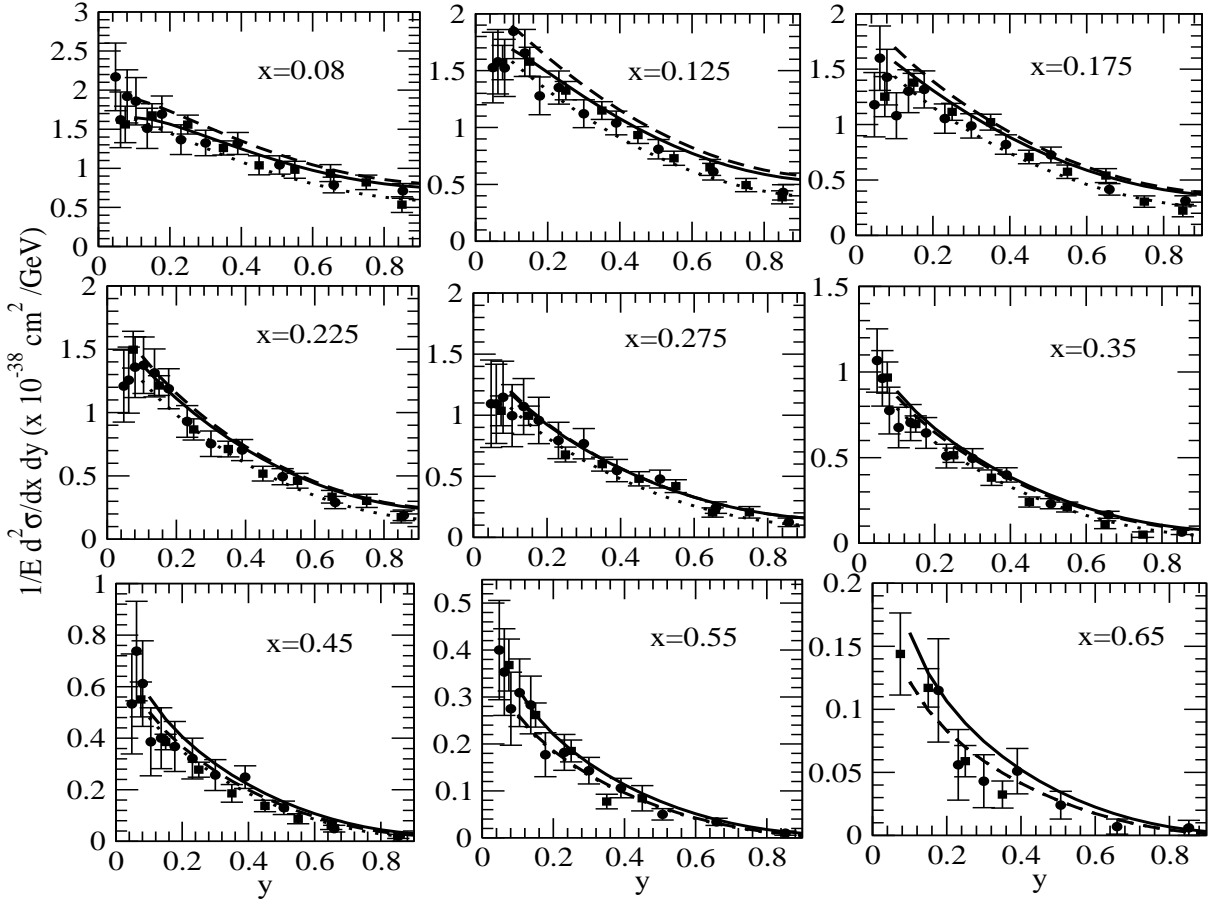}
\caption{$\frac{1}{E}\frac{d^2\sigma}{dxdy}$ vs y at different x for $\bar\nu{_\mu}$($E_{\bar\nu{_\mu}}=150$ GeV) induced reaction in $^{56}$Fe.
 Dotted (Dashed) line is the base (total) calculation at LO. Solid line is the full calculation at NLO. NuTeV~\cite{Tzanov} data have been shown by the squares and CDHSW~\cite{Berge} data by the solid circles.}
\label{antineutrinoiron150gev}
\end{figure}
\begin{figure}
\includegraphics[scale=0.8]{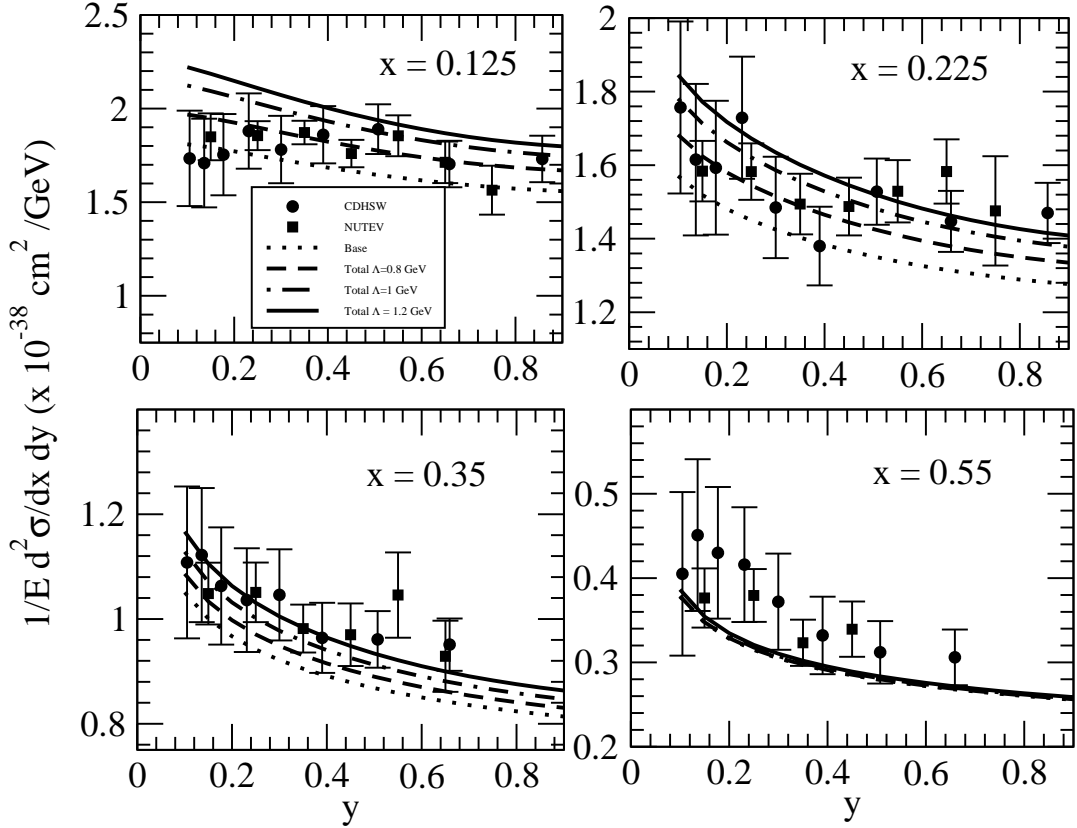}
\caption{$\frac{1}{E}\frac{d^2\sigma}{dxdy}$ vs y at different x for $\nu_\mu$ induced reaction 
in $^{56}$Fe at $E_{\nu_\mu}=65$ GeV using CTEQ~\cite{cteq} PDF at LO. Full model: dashed line with $\Lambda$, $\Lambda_\rho$ = 0.8 GeV, dashed-dotted line with $\Lambda$, $\Lambda_\rho$ = 1 GeV and 
solid line: $\Lambda$, $\Lambda_\rho$ = 1.2 GeV;
 Full model without  pion, rho and shadowing: dotted line. 
NuTeV~\cite{Tzanov} data have been shown by the squares and CDHSW~\cite{Berge} data by the solid circles.}
\label{lambdavariation}
\end{figure}
\begin{figure}
\includegraphics[width=16cm,height=12cm]{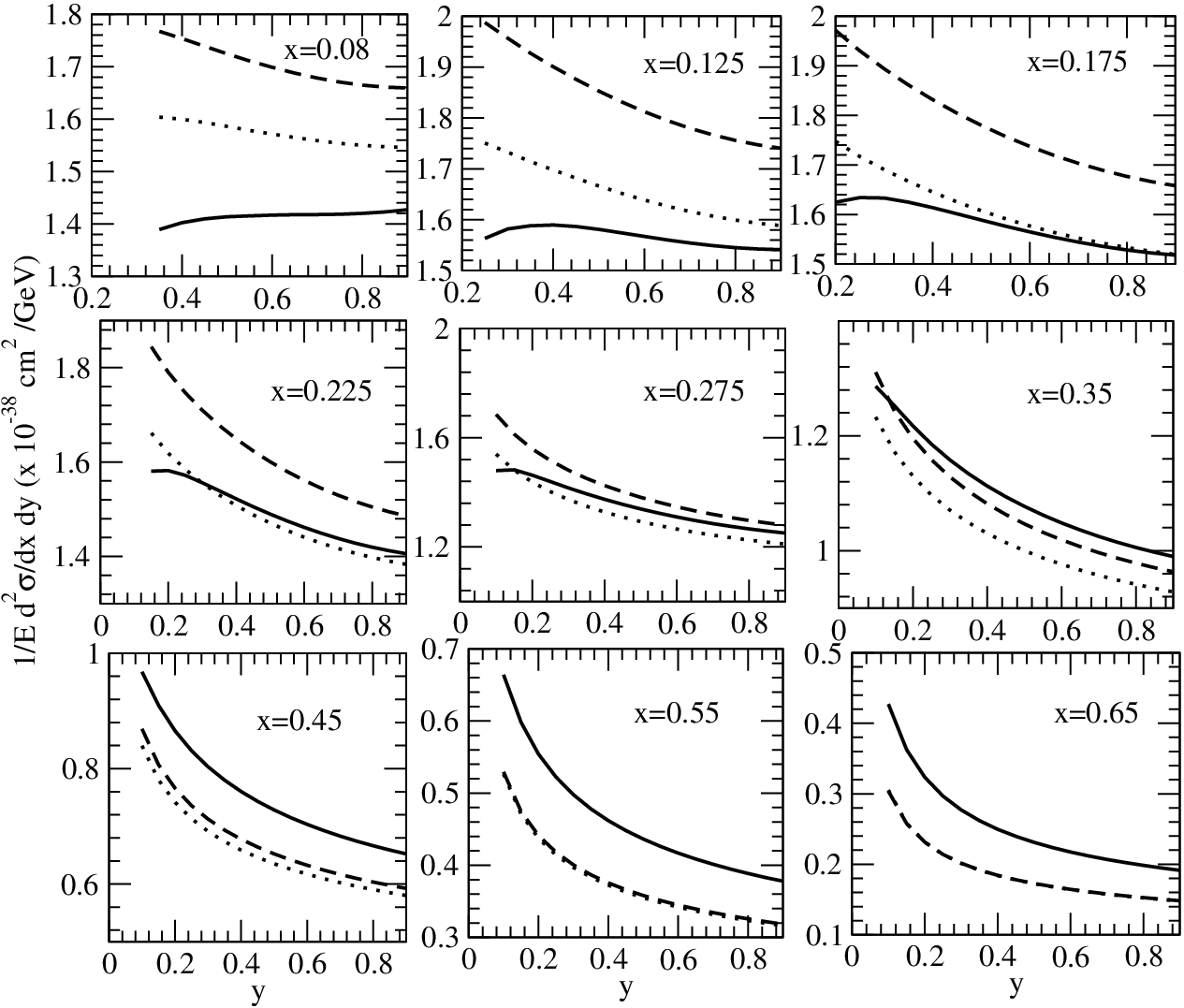}
\caption{$\frac{1}{E}\frac{d^2\sigma}{dxdy}$ vs y at different x for $\nu_\mu$ ($E_{\nu_\mu}=20$ GeV) induced reaction in $^{12}$C. Dotted (Dashed) line is the base (total) calculation at LO. Solid line 
is the result with the full calculation at NLO.}
\label{neutrinocarbon20gev}
\end{figure}
\begin{figure}
\includegraphics[width=16cm,height=12cm]{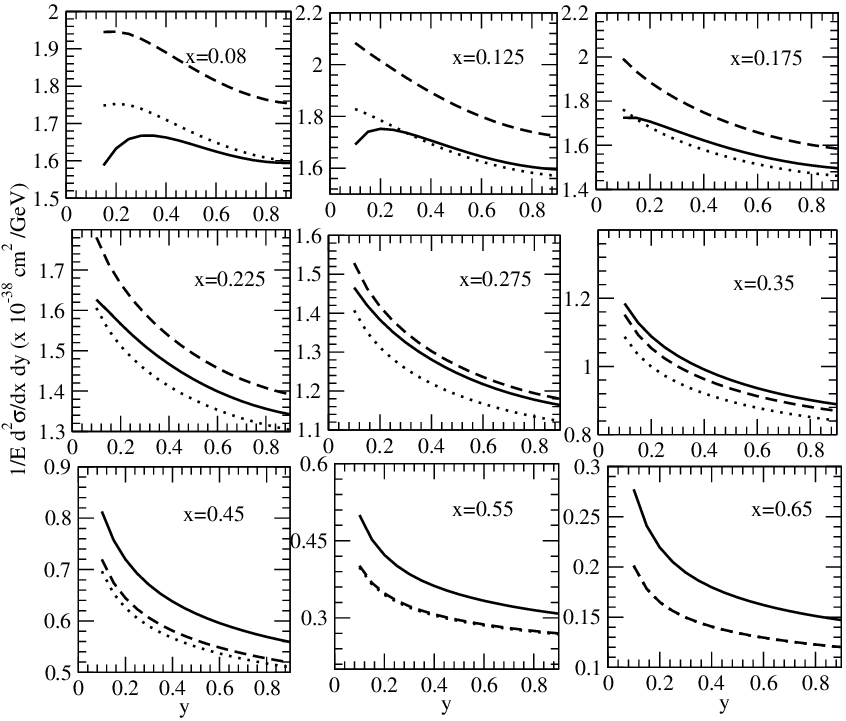}
\caption{$\frac{1}{E}\frac{d^2\sigma}{dxdy}$ vs y at different x for $\nu_\mu$  ($E_{\nu_\mu}=60$ GeV) induced reaction in $^{12}$C. Dotted (Dashed) line is the base (total) calculation at LO. Solid line 
is the result with the full calculation at NLO. }
\label{neutrinocarbon60gev}
\end{figure}
\begin{figure}
\includegraphics[width=16cm,height=12cm]{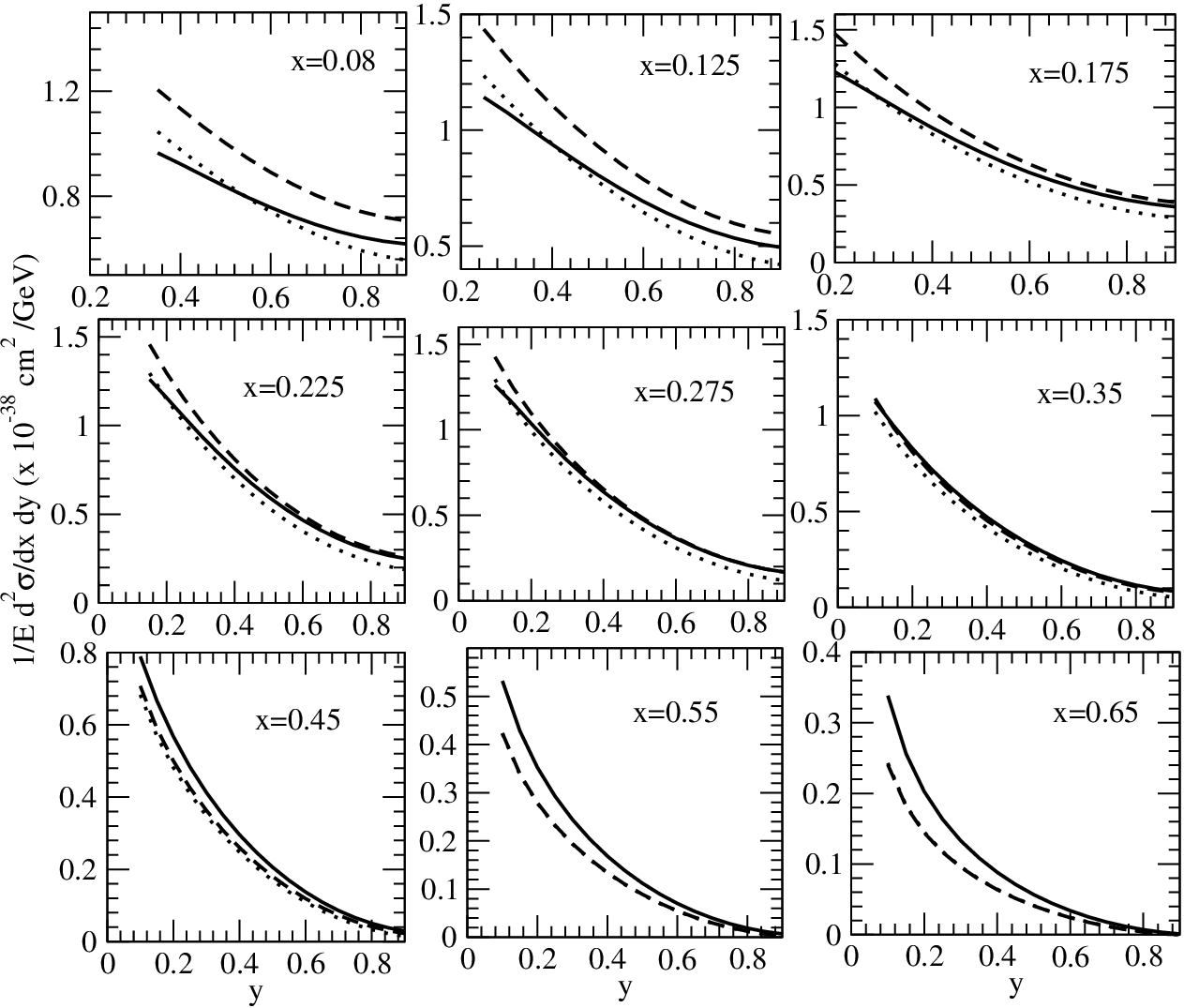}
\caption{ $\frac{1}{E}\frac{d^2\sigma}{dxdy}$ vs y at different x for $\bar\nu{_\mu}$ ($E_{\bar\nu{_\mu}}=20$ GeV) induced reaction in $^{12}$C. 
Dotted (Dashed) line is the base (total) calculation at LO.
 Solid line 
is the result with the full calculation at NLO.}
\label{antineutrinocarbon20gev}
\end{figure}
\begin{figure}
\includegraphics[width=16cm,height=12cm]{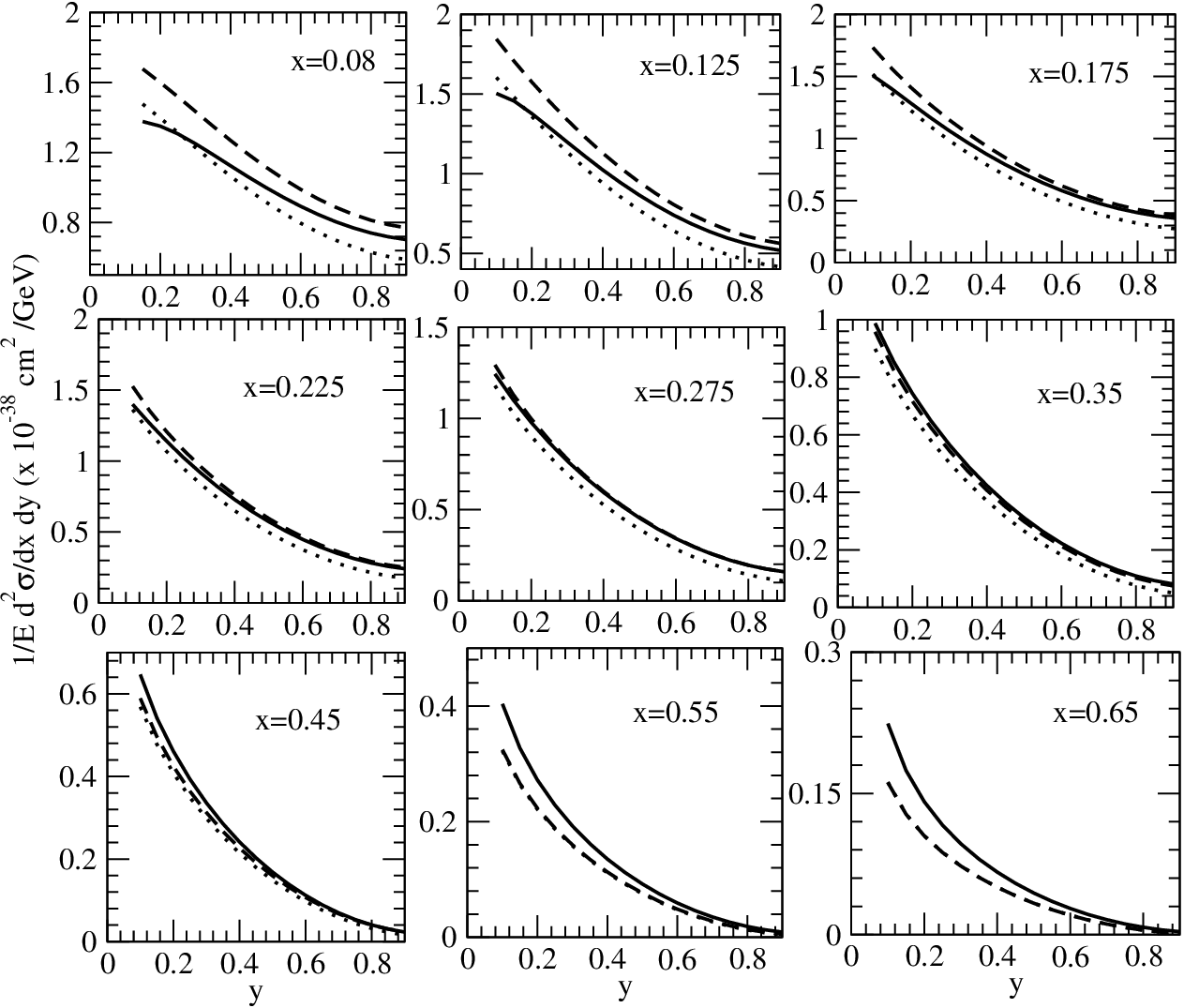}
\caption{$\frac{1}{E}\frac{d^2\sigma}{dxdy}$ vs y at different x for $\bar\nu{_\mu}$ ($E_{\bar\nu{_\mu}}=60$ GeV) induced reaction in $^{12}$C. Dotted (Dashed) line is the base (total) 
calculation at LO. Solid line 
is the result with the full calculation at NLO.}
\label{antineutrinocarbon60gev}
\end{figure}
\section{Results and Discussion}\label{Sec:Results}
In this section we present and discuss the results of our numerical calculations. In the local density approximation, as the cross section is expressed in terms of the nuclear density, for 
the calculations we have used the density for $^{12}$C nucleus to be a harmonic
 oscillator density and for $^{56}$Fe nucleus is taken to be a two Fermi parameter distribution and the density parameters are taken from Ref~\cite{Vries}. 

Using Eqs.(\ref{f2Anuclei}) and (\ref{f3Anuclei}), we have calculated the $F^A_2$ and $F^A_3$ structure functions in the iron nucleus with target mass correction (TMC) and 
CTEQ6.6 parton distribution functions (PDFs) at LO~\cite{cteq}. We call this as our base (Base) result. Thereafter we include pion and rho cloud contributions in $F^A_2$ and the 
shadowing corrections in $F^A_2$ and $F^A_3$, which we call as our full calculation (Total). In Figs.(\ref{fig:f2}) and (\ref{fig:f3}), we have shown these numerical results 
along with the experimental data of NuTeV~\cite{Tzanov} and CDHSW~\cite{Berge} for a wide range of x and $Q^2$. The effect of shadowing is
 about 3-5$\%$ at x=0.1, Q$^2$=1-5GeV$^2$ and 1-2$\%$ at x=0.2, Q$^2$=1-5GeV$^2$ which dies out with the increase in x and Q$^2$.
 In the case of $F^A_2$ there are pion and rho cloud contributions. The pion contribution is very dominant in comparison to the rho contribution. Pion contribution 
is significant in the region of $0.1~<~x~<~0.4$. Thus, we find that the shadowing corrections seem to be negligible as compared to 
the other nuclear effects. It is the meson cloud contribution which is dominant at low and intermediate x for $F_2$. In these figures we also show the results 
of our full calculation at NLO. We find that the results at NLO are
 in better agreement with the experimental observations. 

Recently, we have studied the effect of nuclear medium on the electromagnetic nuclear structure function $F_2(x,Q^2)$ in nuclei using the same model as mentioned in Section-{\ref{Sec:Medium-effect}}. We have obtained the ratio R(x,Q$^2$)=$\frac{2F_{2}^{A}}{AF_2^D}$ (A= $^{9}$Be, $^{12}$C, 
$^{40}$Ca, $^{56}$Fe) and compared our results~\cite{Sajjad1} with the 
recent JLab results of Ref.~\cite{Seely} as well as with some of the older experiments~\cite{Gomez:SLAC}.
The deuteron structure functions have been calculated using the same formulae as in Eqs.(\ref{f2Anuclei}) and (\ref{f3Anuclei})  
but performing the convolution with the deuteron wave function squared instead of the spectral function. 
See Ref.\cite{Kulagin} for full details about the structure functions of deuteron. We have used the parametrization given in Ref. \cite{Lacombe:1981eg} 
for the deuteron wave function of the Paris N-N potential.
We found that the results agree with those of JLab~\cite{Seely}. 

To understand this ratio in the weak sector we have studied the ratio of structure functions R(x,$Q^2$)=$\frac{2F_{i}^{A}}{AF_i^D}$(i=2,3) in the neutrino/antineutrino induced deep inelastic scattering.
This is important as there are several groups~\cite{Hirai, hirai, eskola, Bodek1, Eskola, Kovarik, Kovarik1, Schienbein1} who have phenomenologically studied the nuclear effects in parton 
distribution functions (PDFs). The main differences in their studies are the choice of the experimental data sets and the parametrization of the PDFs at the input level besides some other 
minor differences. In most of these studies 
the experimental data have been taken from the charged lepton nucleus (l$^\pm$A) scattering and the Drell-Yan(DY) data. A few of them also include neutrino scattering data ($\nu$A) in the parametrization
 of nuclear PDFs for the analysis of deep inelastic neutrino/antineutrino cross sections in nuclei. The reliability of nuclear correction factor for the weak interaction induced processes obtained from 
the l$^\pm$A+DY data may be questioned or how good would be the description if one also combines the $\nu$A data. Recently Kovarik et al.~\cite{Kovarik1} have phenomenologically studied nuclear correction factor by taking two 
data sets, one l$^\pm$A+DY data and the other set of $\nu$A  data in iron from the NuTeV measurements and observed that the nuclear effects are 
different particularly at low and intermediate x. Here in the present work we have studied the nuclear effects in the ratio for $\frac{2F_{i}^{A}}{AF_i^D}$(i=2,3) in iron at $Q^2=5$ GeV$^2$ and the results
 are shown in Figs.\ref{fig:f2ratio} and \ref{fig:f3ratio}. Here, we have also shown the results obtained from the phenomenological studies of Tzanov et al.~\cite{Tzanov}, Hirai et al.~\cite{hirai}, 
Eskola et al.~\cite{Eskola} and Schienbein et al.~\cite{Schienbein1}. We find that our results for the ratio of R(x,Q$^2$)=$\frac{2F_{2}^{A}}{AF_2^D}$ are similar to what we have obtained in the case of 
electromagnetic interaction~\cite{Sajjad1}, while the ratio R(x,$Q^2$)=$\frac{2F_{3}^{A}}{AF_3^D}$ is different in nature. It may be seen that the results of the different phenomenological studies differ 
among themselves as well as to our results. Whereas in most~\cite{Tzanov, hirai} of the phenomenological analyses the nuclear correction factor in $F_2$ and $F_3$ are taken to be the same 
we are finding it to be different. Although the nuclear effects like Fermi motion and binding corrections
 are the same in $F_2$ and $F_3$ which have been incorporated by means of the use of a spectral function obtained for nuclear matter and implemented in nuclei using the 
local density approximation, the differences in the results of $F_2$ and $F_3$ in our model are due to the fact that in the $F_2$ 
structure function there are meson cloud contributions whereas in $F_3$ this is absent, there is a different target mass correction(TMC) and a different kinematical 
factor as can be seen from Eqs.(\ref{f2Anuclei}) and (\ref{f3Anuclei}). We have observed that the effect of meson clouds are large at low and intermediate x. There is an almost negligible shadowing
 correction. Therefore, we conclude that it is not appropriate to take the same correction factor for the $F_2$ and $F_3$ nuclear structure functions.

In Figs.(\ref{neutrinoiron65gev}) and (\ref{neutrinoiron150gev}), we have shown the results for $\frac{1}{E}\frac{d^2\sigma}{dxdy}$ in $^{56}$Fe at $E_{\nu_\mu}$=65 GeV and 150 GeV respectively. 
The calculations
 for the the double differential cross sections have been performed for $Q^2~>~1GeV^2$. Similarly in Figs.(\ref{antineutrinoiron65gev}) and 
(\ref{antineutrinoiron150gev}), we have shown the results for $\frac{1}{E}\frac{d^2\sigma}{dxdy}$ induced by antineutrinos in $^{56}$Fe at $E_{\bar\nu_\mu}$=65 GeV and 150 GeV respectively. 
We find that the results of the full calculations at NLO are in general in good agreement with the experimental observations of CDHSW~\cite{Berge} and NuTeV~\cite{Tzanov} collaborations.

In the present model, for the pion and rho mesons contributions the expressions for which are given in Eqs.(\ref{F2pion}) and (\ref{F2rho}) respectively, the expression includes 
pion and rho meson self energies~\cite{Marco} which has been earlier quite successfully used in the calculations of pion, electron and photon induced reactions in nuclei. It has some uncertainties such as the 
specific form of the spin-isospin interaction, specially for the $\rho$ meson. To understand the effect of the variation in the parameters of the pion and rho meson self energies, $\Lambda$ and $\Lambda_\rho$ 
respectively, on the differential scattering cross section we have plotted $\frac{1}{E}\frac{d^2\sigma}{dxdy}$ in Fig.(\ref{lambdavariation}) by taking $\Lambda$, $\Lambda_\rho$=0.8GeV, 
1.0GeV and 1.2GeV. We find that a 20$\%$ variation in $\Lambda$'s, results in a change of 4-6$\%$ in the cross section at low x which decreases to 2-3$\%$ around x=0.4-0.5 and after that it dies out. 
To observe the effect of non-isoscalarity, we have also studied (not shown) isoscalarity vs non-isoscalarity corrections and found that the non-isoscalar correction is around 2-3$\%$. 

Figs.(\ref{neutrinocarbon20gev}) and (\ref{neutrinocarbon60gev}) are the results for $\frac{1}{E}\frac{d^2\sigma}{dxdy}$ in $^{12}$C induced by neutrinos at $E_{\nu_\mu}$=20 GeV and 60 GeV respectively and 
Figs.(\ref {antineutrinocarbon20gev}) and (\ref{antineutrinocarbon60gev}) are the corresponding results in $^{12}$C induced by antineutrinos. The 
results in carbon will be useful in the analysis of data by the NOMAD~\cite{nomad} collaboration as well as the proposed NuSOnG \cite{nusong,nusong1}. NOMAD~\cite{nomad1} experiment is primarily meant to
 measure the neutrino/antineutrino cross sections with better precision and constrain the nuclear models. Therefore, our study of the nuclear medium effects would be a good test when the data will be available.
\section{Conclusions}\label{Sec:Conclusion}
To summarize our results, we have studied nuclear effects in the structure functions $F^A_{2} (x, Q^2)$ and $F^A_{3} (x, Q^2)$ in the carbon and iron nuclei using a many body theory to describe 
the spectral function of the nucleon in the nuclear medium for all $Q^2$. Local density approximation has been used to apply the results for the finite nuclei. The use of spectral 
function is to incorporate Fermi motion and binding effects. We have used CTEQ~\cite{cteq} PDFs in the numerical evaluation. Target mass correction (TMC) has been considered.
We have taken the effects of mesonic degrees of freedom, shadowing, anti-shadowing in the calculation of $F_2^A$ and shadowing and anti-shadowing effects in the calculation of $F^A_{3}$. 
We have found that the mesonic cloud (basically pion) gives an important contribution to the cross section. These numerical
results have been compared with the experimental observations of CDHSW~\cite{Berge} and NuTeV~\cite{Tzanov} collaborations. Using these structure functions we obtained differential scattering cross section 
for iron and carbon nuclei and compared the results in iron from the experimentally observed values obtained by CDHSW~\cite{Berge} and NuTeV~\cite{Tzanov} collaborations.  We also find that the effect of nuclear medium is also quite important even in the case of deep inelastic scattering 
and the ratio of the structure functions in nuclei to deuteron or free nucleon is different in $F^A_{2} (x, Q^2)$ and $F^A_{3} (x, Q^2)$.  
\section{Acknowledgment}
\begin{acknowledgments}
This research was supported by DGI
and FEDER funds, under contracts FIS2008-01143/FIS, FIS2006-03438, and the Spanish Consolider-Ingenio 2010
Programme CPAN (CSD2007-00042), 
by Generalitat Valenciana contract PROMETEO/2009/0090 and by the EU
HadronPhysics2 project, grant agreement n. 227431. 
I.R.S. acknowledges support from the Ministerio de Educaci´on. M. S. A. wishes to acknowledge the financial support from the University of Valencia and Aligarh Muslim University under the academic
 exchange program and also to the DST, Government of India for the financial support under the grant SR/S2/HEP-0001/2008. H. H. acknowledges Maulana Azad National Fellowship. 
\end{acknowledgments}

\end{document}